\documentclass[aps,pra,reprint]{revtex4-1}

\pdfoutput=1

\usepackage{amsmath}%
\usepackage{amsfonts}%
\usepackage{amssymb}%
\usepackage{graphicx}

\begin{document}

\title{Insensitivity of Ion Motional Heating Rate to Trap Material over a Large Temperature Range}

\author{J. Chiaverini}
\email[]{john.chiaverini@ll.mit.edu}
\affiliation{MIT Lincoln Laboratory, RF \& Quantum Systems Group, 244 Wood St, Lexington, MA 02420 USA}

\author{J. M. Sage}
\email[]{jsage@ll.mit.edu}
\affiliation{MIT Lincoln Laboratory, RF \& Quantum Systems Group, 244 Wood St, Lexington, MA 02420 USA}

\date{\today}

\begin{abstract}
We present measurements of trapped-ion motional-state heating rates in niobium and gold surface-electrode ion traps over a range of trap-electrode temperatures from approximately $4$~K to room temperature ($295$~K) in a single apparatus.  Using the sideband-ratio technique after resolved-sideband cooling of single ions to the motional ground state, we find low-temperature heating rates more than two orders of magnitude below the room-temperature values and approximately equal to the lowest measured heating rates in similarly-sized cryogenic traps.  We find similar behavior in the two very different electrode materials, suggesting that the anomalous heating process is dominated by non-material-specific surface contaminants.  Through precise control of the temperature of cryopumping surfaces, we also identify conditions under which elastic collisions with the background gas can lead to an apparent steady heating rate, despite rare collisions.  
\end{abstract}


\maketitle 

A system of trapped atomic ions addressed with electromagnetic fields is a promising candidate for large-scale quantum processing~\cite{RevModPhys.75.281}, but to achieve fault-tolerance, multi-qubit gate fidelity must be increased~\cite{Blatt:Wine:Nat08}.  As all multi-qubit ion gates demonstrated to date rely on the Coulomb interaction to manipulate internal states via external, motional states in a harmonic potential, electric field noise near the trap-frequency can lead directly to gate error.  Indeed, motional-state heating~\cite{Turchette:d4heating:PRA:00} currently limits two-qubit ion gate errors at or above the $10^{-3}$ level~\cite{PhysRevLett.108.260503,PhysRevLett.110.263002}.  This electric-field noise appears to originate from (typically highly conductive) trap-electrode surfaces, and it seems to be much larger than the expected Johnson (thermal) noise.  The origin of this noise is currently unknown, but it appears to be a thermally activated process, as it is greatly reduced at low trap-electrode temperature~\cite{PhysRevLett.97.103007,Chaung:CryoHeatRates:PRL08}.  It is also greatly reduced via in-vacuum ion-milling of the electrodes~\cite{PhysRevLett.109.103001,1307.7194}, suggesting that surface effects are the primary source.

Here, using a specialized cryogenic UHV apparatus, we present ion heating rate measurements over a more-complete temperature range than has been previously demonstrated.  We present temperature-dependent heating rates in niobium traps for the first time, and we extend the range of measurements in gold traps beyond previously published work~\cite{MIT:HeatingvsT:PRL:08}, measuring heating rates up to room temperature in both materials in the same system.  We also explore motional decoherence caused by background gas collisions in order to rule out the possibility in our experiments.  It is hoped that with these and further temperature- and material-dependent data, the mechanism of anomalous ion heating may be determined and the error from heating subsequently reduced, possibly through development of a materials or (ex situ) processing solution.

Using an apparatus described previously~\cite{PhysRevA.86.013417}, we load individual $^{88}\rm{Sr}^{+}$ ions into segmented, single-zone, surface-electrode linear Paul traps~\cite{NIST:SET:QIC:05} $50~\mu$m from the electrode surface.  The traps are fabricated from sputtered niobium ($2$~$\mu$m thick), or thermally evaporated gold ($500$~nm thick), on a sapphire substrate.  Chips are patterned via standard photolithography and wet (Au) or dry (Nb) etching with minimum gaps between electrodes near the ion location of $4$--$6$~$\mu$m.  A radio frequency (RF) potential of approximately $35$~V amplitude at $27$~MHz is applied using a helical resonator to produce radial secular frequencies of $4$--$5$~MHz.  DC potentials up to $25$~V are applied to the control electrodes to produce an axial trap frequency of $1.32$~MHz using a commercial, multi-channel, computer-controlled digital-to-analog converter~\footnote{National Instruments PXI-6723.} (DAC) followed by home-built DC amplifiers (gain $\sim\nobreak4$).  From potential simulations we calculate a trap depth in this configuration of $\sim\nobreak15$~meV.  To reduce electric field noise near the trap frequency, the control electrode lines are filtered close to the trap with single-stage low-pass RC~filters (cutoff frequency $\sim\nobreak16$~kHz) at low temperature and outside the chamber with multi-stage low-pass filters~\footnote{Kiwa Electronics, Kasson, MN 55944 USA} (cutoff frequency $\sim\nobreak50$~kHz; we measure $>95$~dB rejection at $1.32$~MHz). 

The trap chip is attached to a small copper block (the trap stage) which contains a resistive heater.  The trap stage is connected via a weak thermal link to the low-temperature (LT) stage of a vibration-isolated cryocooler.  The dominant thermal conductance between the trap stage and the LT stage is through approximately $50$~gold wire bonds and allows a large temperature gradient between the two stages.  At base temperature, approximately $3.5$~K at the LT stage, the trap chip gets below $5$~K, as measured by a calibrated diode temperature sensor mounted on the chip, with RF potential applied and cooling laser beams on.  The chip can be heated via the trap stage to room temperature while maintaining the LT stage at a temperature of $17$~K or lower, sufficient for UHV operation.  The trap chip, trap stage, and LT stage are surrounded by a thermal radiation shield held at a temperature of approximately~$55$~K.

\begin{figure}[tbp]
\includegraphics[width=\columnwidth]{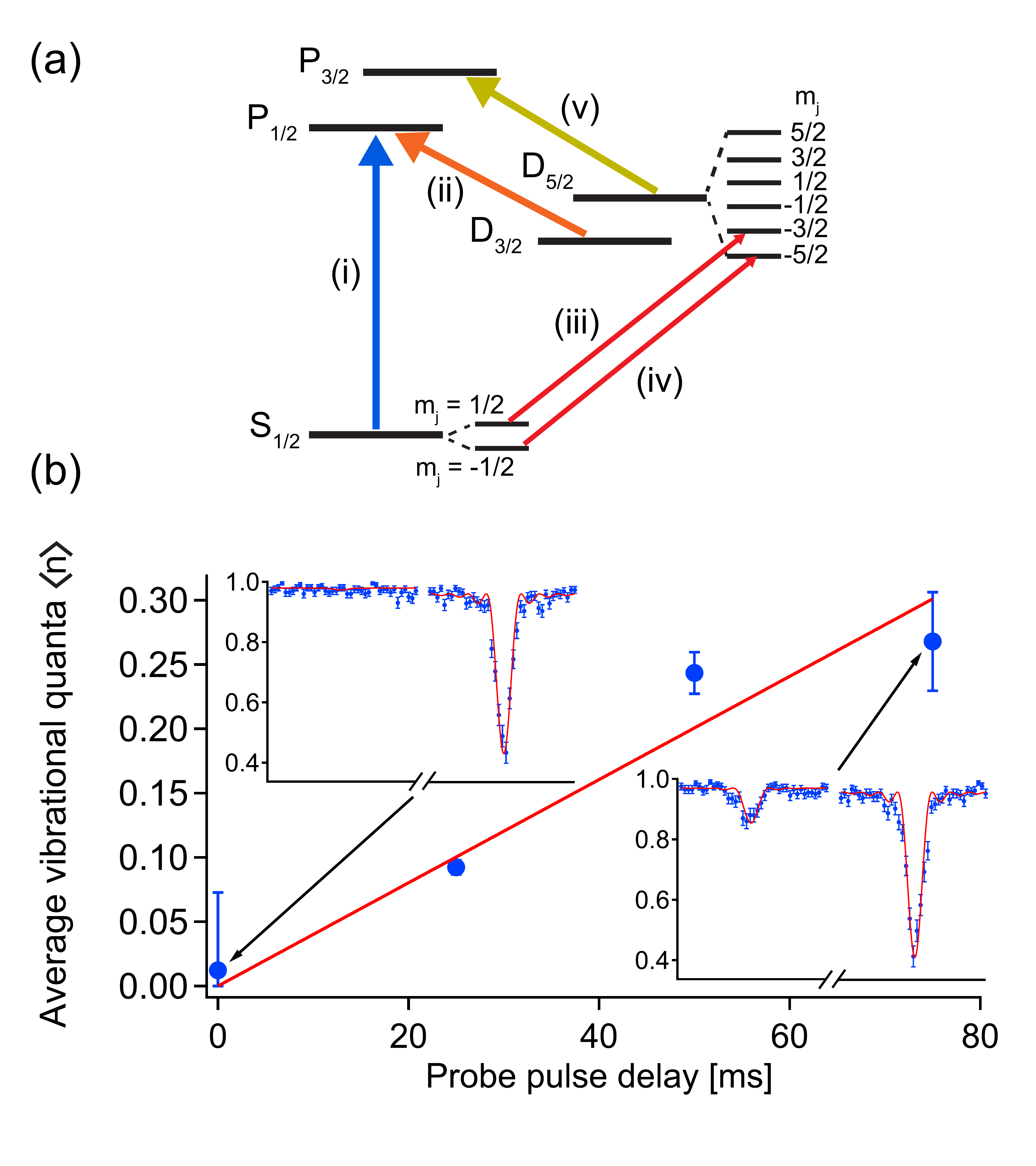}
\caption{(Color online) Ion levels and motional decoherence measurement.  (a) Electronic level diagram of $^{88}$Sr$^{+}$ including relevant transitions and Zeeman sub-levels. Transitions (i) and (ii) are used for Doppler cooling and repumping to keep the ion in the cooling cycle, respectively.  These transitions are also used for ion state detection.  The others are used for (iii) optical pumping (OP), (iv) sideband cooling (SC) and probing, and (v) quenching during OP and SC and before each experiment.  (b) Example of data from a heating rate measurement in a niobium trap at $3.8$~K using the sideband spectroscopy method.  The red and blue sidebands of the $S_{1/2}$--$D_{5/2}$ transition (insets at particular probe delays) are fit to a Rabi line shape to extract their amplitudes, which are used to calculate $\langle n \rangle$ as a function of probe delay (main figure) after cooling to the ground state.  A linear fit to these points provides the heating rate.  Insets show $S$-state probability as a function of probe frequency near the sidebands.  Errors depicted in the insets are statistical including projection noise and are propagated through the sideband fits to produce errors in the main figure points.\label{figure:nbarvt}}

\end{figure}

We load single ions into the trap from a pre-cooled source of magneto-optically trapped neutral strontium atoms (MOT).  The MOT is loaded from an oven which has no direct line of sight to the trap chip~\cite{PhysRevA.86.013417}.  The ions are cooled to approximately the Doppler limit ($\sim\nobreak0.5$~mK) by excitation on a strong optical transition ($S_{1/2}$--$P_{1/2}$) and simultaneous repumping from the long-lived $D_{3/2}$ level to the $P_{1/2}$ level (see Fig.~\ref{figure:nbarvt}a for a description of the relevant ion energy levels).  The kinetic energy of the ions is further reduced by performing pulsed resolved sideband cooling~\cite{PhysRevLett.75.4011} using a narrowed (linewidth $<5$~kHz) 674~nm laser red-detuned by the trap frequency from the $|S_{1/2},m_{j}=-1/2\rangle$ $\rightarrow$ $|D_{5/2},m_{j}=-5/2\rangle$ transition, in combination with quenching light near $1033$~nm to couple the $D_{5/2}$ to the $P_{3/2}$ and return the ion to the $S$ state.  Sideband cooling pulses are interspersed with optical pumping (OP) pulses resonant with the $|S_{1/2},m_{j}=1/2\rangle$ $\rightarrow$ $|D_{5/2},m_{j}=-3/2\rangle$ transition.  A small magnetic field ($\sim\nobreak6.5$~G) is applied along the direction perpendicular to the trap axis and parallel to the trap surface to lift the Zeeman degeneracy of these levels for spectroscopic addressing of the $S_{1/2}$--$D_{5/2}$ transition and to prevent the formation of dark states in the $D_{3/2}$ manifold.  Using this technique we achieve $>99$\% occupation of the axial mode vibrational ground state (the $n=0$ state) for the lowest heating rates~\footnote{For the higher heating rates seen near room temperature, we typically achieve $\protect\langle n\protect\rangle<0.25$ measured after sideband cooling and probing, limited by our achievable cooling rate and probe time at these trap frequencies.}.

For heating rate measurements, each experiment begins with Doppler and sideband cooling with OP to prepare the ion in the motional ground state of the $|S_{1/2},m_{j}=-1/2\rangle$ level.  After a variable delay time and probe pulse, the ion state is detected via the presence (absence) of fluorescence if the ion is in the $S_{1/2}$ ($D_{5/2}$) state during application of light nearly resonant with the $S_{1/2}$--$P_{1/2}$ transition.  In multiple experiments at a fixed delay time, we probe around the red and blue sidebands of the $|S_{1/2},m_{j}=-1/2\rangle$ $\rightarrow$ $|D_{5/2},m_{j}=-5/2\rangle$ transition.  We use the red-to-blue sideband amplitude ratio $r$ to determine the ion's average vibrational mode occupation~\cite{PhysRevLett.75.4011} $\langle n\rangle=r/\left( 1-r \right)$ for that delay, and then we vary the delay to determine the heating rate at a particular temperature.  We perform $200$--$400$ experiments at each probe frequency, delay, and temperature.  A typical measurement of the heating rate is shown in Fig.~\ref{figure:nbarvt}b.  See Appendix~\ref{app_methods} for more details on experimental methods.

\begin{figure}[tbp]
\includegraphics[width=\columnwidth]{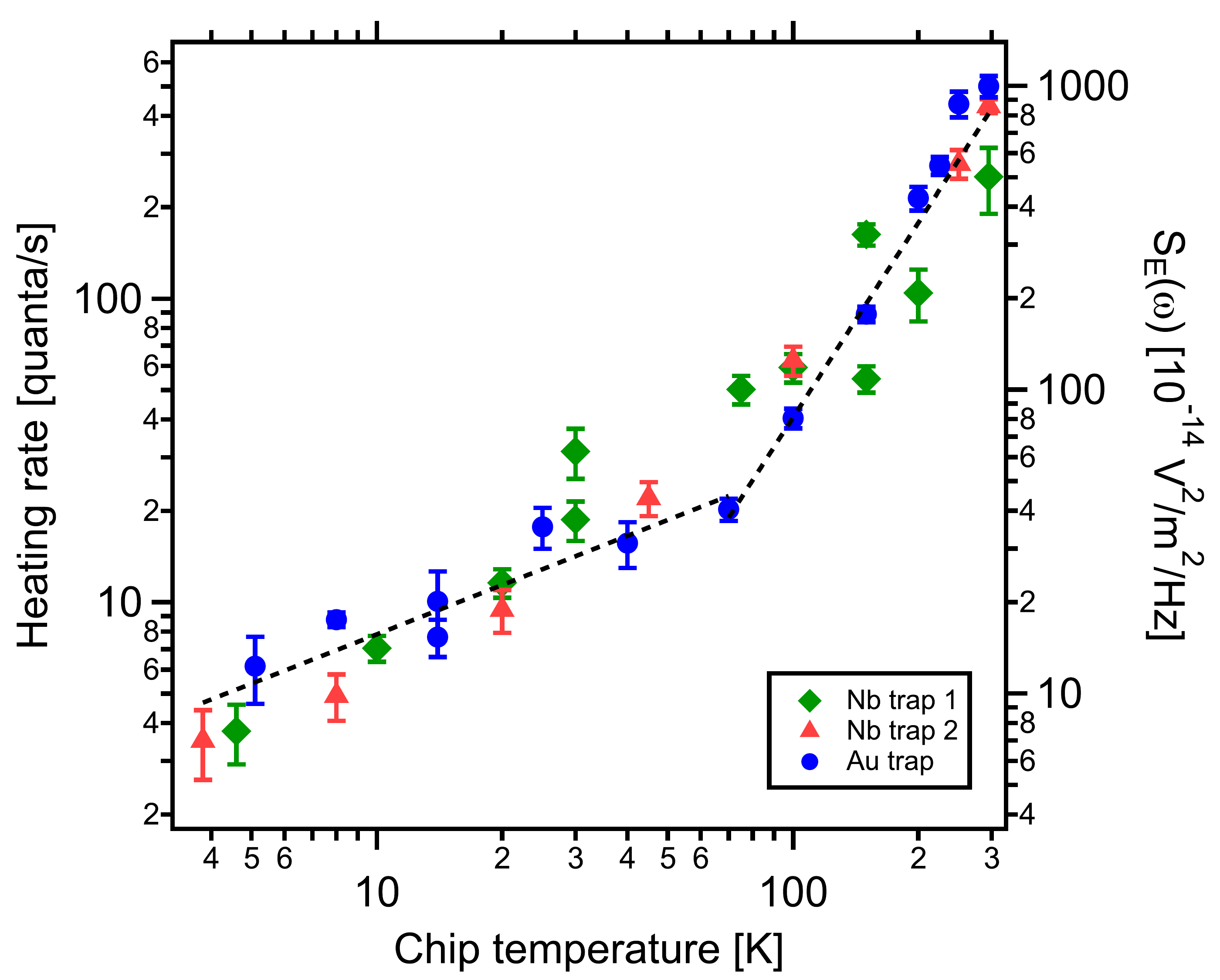}
\caption{(Color online) Motional heating rate of the $1.32$~MHz axial mode versus trap temperature for two niobium traps and a gold trap.  The right axis is translated into electric field noise spectral density via $S_{E}\left( \omega \right)=\frac{4 m \hbar \omega}{q^2}\frac{d \langle n \rangle}{dt}$.  Here $\omega=2\pi\times f$ is the angular trap frequency, $q$ is the ion charge, and $m$ is the ion mass.  All error bars are statistical based on errors propagated through fits of the sideband spectra (cf. Fig.~\ref{figure:nbarvt}).  The dotted lines are fits to power laws in temperature to all data; see text for details.  For $70$~K and above, the scaling exponent is~2.13(5), and below this temperature, the exponent is~0.54(4).\label{figure:HRvT}}
\end{figure}

In Fig.~\ref{figure:HRvT} we present the ion heating rate of the $1.32$~MHz axial vibration mode as a function of chip temperature for two niobium traps and one gold trap.  The heating rate drops smoothly below room temperature, with a factor of approximately~$100$ between room temperature and $4$~K, and it is quite consistent over two orders of magnitude in both temperature and heating rate between the materials.   Rates at the lowest chip temperatures are among the lowest ever measured for traps smaller than a $75$~$\mu$m ion-electrode distance (see also Fig.~\ref{figure:omegaSEvd} below).  Room temperature heating rates are consistent with, or lower than, those seen by others in microfabricated traps of similar size.  It should be noted that the niobium traps superconduct below $9.2(1)$~K as determined via a four-wire resistance measurement of an on-chip meander resistor co-fabricated in the trap film~\footnote{The transition temperature determined in situ is equal within error to that measured separately using another niobium trap chip in a probe lowered into a liquid helium dewar.}.  Consistent with previous results~\cite{wang:244102}, we see no significant difference in the heating rate just below and just above the superconducting transition.

The observation of overall temperature dependence very similar in the Nb and Au traps studied here, each with very different material properties, suggests that anomalous heating is due primarily to adsorbates or contaminants on the surface, rather than material-specific surface properties.  Niobium oxidizes readily, while gold is somewhat noble, so their similar behavior in these experiments would appear to place constraints on noise origins that are strongly dependent on native insulating oxide patches that may contain mobile charges.  Additionally, contact potentials due to varying work functions across a material surface have been suggested as the origin of this noise~\cite{NIST:ExpIssueswithIons:JresNIST:98,PhysRevB.32.7703}, but our results suggest that the temperature dependence of two materials with different work functions and presumably different polycrystallinity is not particularly different.  The trap chips studied here have undergone similar fabrication, cleaning, and installation techniques, however, possibly implicating these processing steps.  This conclusion is consistent with recent demonstrations of reduced ion heating in traps after in vacuo ion milling~\cite{PhysRevLett.109.103001,1307.7194}.

Our vacuum system is not baked; we depend on cryopumping to attain low pressures.  Therefore we do not expect the same hydrocarbon contamination that has been seen in stainless steel chambers subjected to high-temperature bakes~\cite{PhysRevLett.109.103001}.  If such post-chamber-bake contamination is the dominant source of anomalous heating in baked systems, it may be that ex~situ milling followed immediately by installation into a vacuum system not requiring a bake (notably a cryogenic system) could lead to improved heating rates without the more involved accommodation of an in situ milling capability.

As the different traps show similar heating rate temperature dependence over the entire range, and since we suspect the overall trend may be due to non-material-specific surface properties, we determine temperature scaling using the combined data from all three traps.  We note two regions with distinct power law behaviors $\frac{d\langle n\rangle}{dt}\propto T^{\beta}$.  From room temperature to $70$~K our power law fit produces an exponent of~$\beta=2.13(5)$, while below $70$~K the best fit exponent is~$\beta=0.54(4)$.  We suspect the variation between traps is due to so-far uncontrolled day-to-day variability in the measurements or slight, unmarked differences in processing.  The reduction in scaling power at the low end of the temperature range may suggest the approach to a plateau potentially due to temperature-independent heating processes.  Below, we present evidence that similar saturation behavior can be mimicked by another (non-chip-temperature related) process, though we can rule it out in this case.

\begin{figure}[tbp]
\includegraphics[width=\columnwidth]{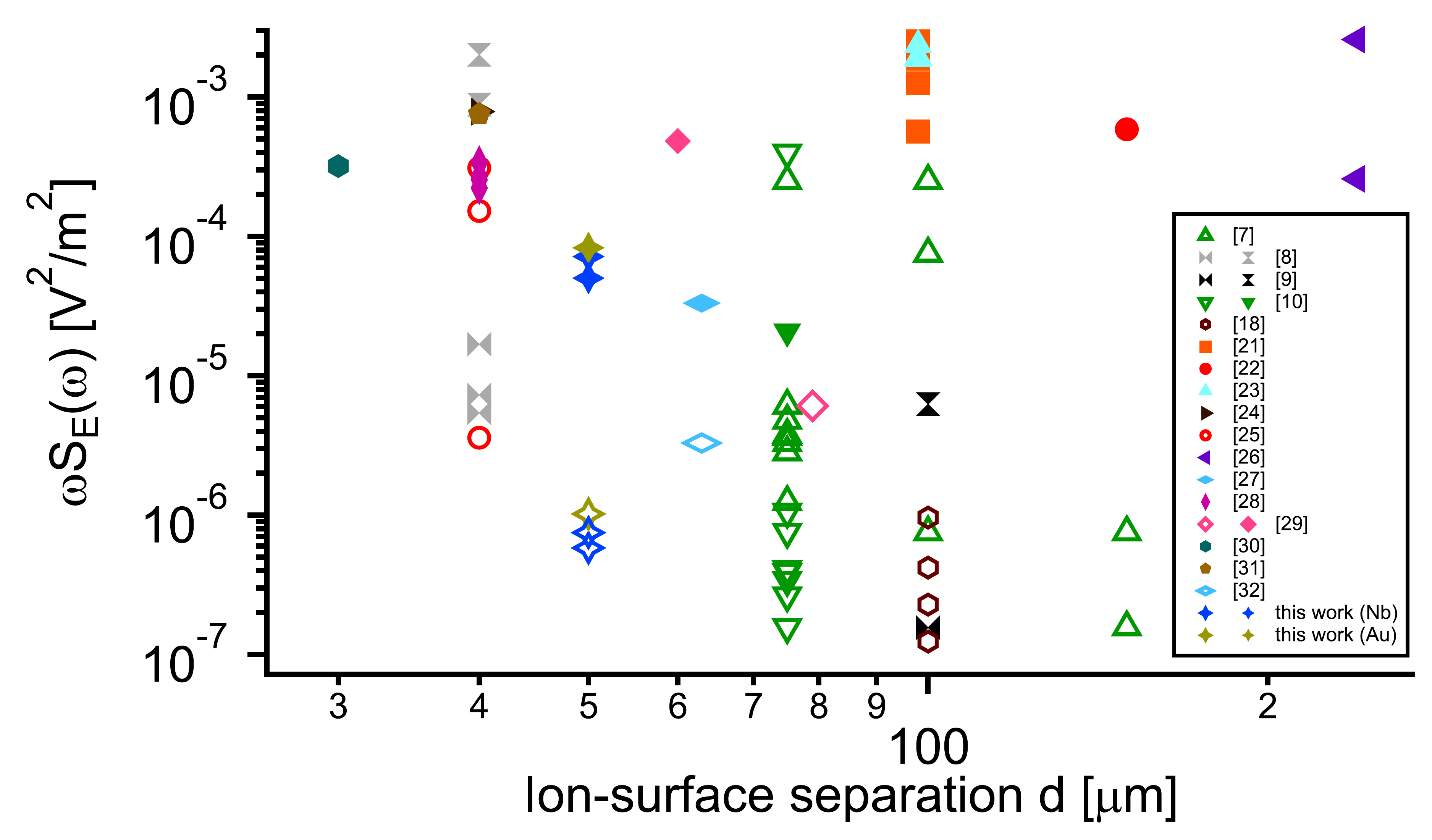}
\caption{(Color online) Electric field noise spectral density, normalized by angular trap frequency, as a function of ion-electrode distance $d$. Measurements in surface traps from several groups are included.  Closed (open) symbols are room-temperature (cryogenic) measurements.  The measurements associated with this work are the blue (gold) four-pointed stars for the niobium (gold); in the case of this work, the open symbols represent data taken at $3.8$--$5.1$~K.  Other groups' cryogenic measurements are taken over a range of $4$--$10$~K (see individual references for details).  Bow-tie symbols represent room temperature data from groups incorporating in situ surface milling: vertical (horizontal) bow ties are for pre-(post-)milling measurements.  References~\cite{1367-2630-13-12-123023,Oxford:TrapDesign:NJPhys:10,Oxford:CaCharging:APB:11,NIST:Bksideloading:APL:09,brown2011coupled,1367-2630-13-1-013032, 1367-2630-14-7-073012,PhysRevA.76.033411,PhysRevLett.109.103001,MIT:HeatingvsT:PRL:08,Chaung:CryoHeatRates:PRL08,SMIT:QIC:2009,NIST:Microwave2qubit:Nat:11, NIST:SET:PRL:06,wang:244102,1307.7194,vittorini:043112} for all plotted data are given in the legend.\label{figure:omegaSEvd}}

\end{figure}

Fig.~\ref{figure:omegaSEvd} shows the present results alongside those from other heating rate studies performed in surface-electrode traps.  Here electric field noise spectral density is shown normalized by trap frequency to better compare noise from disparate experiments.  The low-electrode-temperature noise levels in both the niobium and gold traps measured here are consistent with the lowest levels seen in any other ion trap, assuming that the noise scales as a high inverse power of the ion-electrode separation $d$ ($S_{E}\propto d^{-4}$ is expected for patch-potential-variation-induced noise or random dipole noise~\cite{Turchette:d4heating:PRA:00}).  The Johnson noise expected for our apparatus at $4$~K is $S_{E}\approx1\times10^{-17}$~V$^{2}$/m$^{2}$/Hz, limited by the loss in the filter capacitors~\footnote{The resistance of the electrodes contributes significantly to the Johnson noise only at higher temperatures where the metal resistance becomes comparable to the Th\'{e}venin equivalent resistance of the filters.  For both materials the expected Johnson noise is far below the measured electric field noise over the entire temperature range.}.  This is still more than three orders of magnitude below the noise level corresponding to our lowest heating rates, $S_{E}=7\times10^{-14}$~V$^{2}$/m$^{2}$/Hz.

Our apparatus affords us control of not only the trap temperature, but also the temperature of the coldest part of the vacuum system, the LT stage.  This stage has an additional temperature sensor and heater with a separate control loop from the trap stage.  We have utilized this capability to examine the ion heating rate in a gold trap at fixed electrode temperature while varying the background gas density.  As the LT stage is heated, cryopumping becomes less effective, leading to higher gas density near the ion.  While we cannot directly measure the pressure at the ion location, we can see a proxy of the local density in the trapped-ion heating rate and lifetime.

In the measurements of the gold trap presented above, the LT stage temperature was at $\sim5$~K for trap temperatures of $\sim25$~K.  After taking these measurements, we raised the temperature of the LT stage to $16$~K while maintaining the trap temperature at $25$~K.  Under these conditions, the ion lifetime with cooling light was approximately $10$~s, compared to $\sim\nobreak10$~min with no LT stage heating; due to rapid, automated ion reloading~\cite{PhysRevA.86.013417}, we are able to efficiently measure heating rates despite very short lifetimes.  The heating rate in this case was higher by approximately a factor of $2.5$ ($\sim\nobreak32$~quanta/s compared to $\sim\nobreak13$~quanta/s).  We subsequently turned off the LT-stage heaters and recovered the initial heating rate for the same trap temperature within minutes of the LT reaching its base temperature.  This sequence was repeated, and these heating rates were reproducible.  

Assuming that the apparent extra heating is due to elastic collisions with background gas molecules, we can estimate the local density from  the measured heating rate (see Appendix~\ref{app_collisions} for details; we summarize the results here).  Such elastic collisions impart significant kinetic energy to the ion, heating it after one collision such that the red and blue sidebands will have equal intensity, though not to such an energy that it would be ejected from the trap~\footnote{The maximum expected kinetic energy of a Sr$^{+}$ ion after a collision with a $55$~K H$_{2}$ molecule is $0.4$~meV.  Nitrogen (N$_{2}$) or O$_{2}$ could impart up to $5$~meV, but we do not expect significant amounts of these gases in this case.}.  However, these events happen significantly less often than once per experiment, i.e. $\gamma_{e}t_{D}\ll 1$ where $\gamma_{e}$ is the elastic collision rate and $t_{D}$ is the experiment time including the probe delay, the dominant contribution.  When a few experiments with collision events are averaged with many experiments without a collision, one will measure a red-sideband amplitude proportional to the elastic collision rate and $r\approx\gamma_{e}t_{D}$ in the absence of additional heating.  We estimate the background density that would lead to the measured heating rate due to combined collisional~\cite{NIST:ExpIssueswithIons:JresNIST:98} and anomalous heating to be $1.0(1)\times10^{9}$~cm$^{-3}$~\footnote{For comparison, the background density in a room-temperature system at a pressure of $1\times10^{-10}$~T is approximately $3\times10^{6}$~cm$^{-3}$.}, corresponding to an elastic collision rate of $10$~s$^{-1}$.  Inelastic collisions from the additional background atoms are most likely responsible for the reduced lifetime $\tau_{L}$ at higher background density.  The calculated density leads to an inelastic (Langevin) collision rate~\cite{NIST:ExpIssueswithIons:JresNIST:98} of $0.6$~s$^{-1}$ to $1.6$~s$^{-1}$ for background species of interest.  This is roughly consistent with our ion loss rate under these circumstances ($\tau_{L}^{-1}\sim\nobreak0.1$~s$^{-1}$) considering that the calculated Langevin rate is an upper bound.

Since we expect the ion loss rate to be proportional to the local density, we are confident that the ultimate heating rates measured here (without LT stage heating) are not collision-limited; from the ion lifetimes we estimate the extra heating rate due to collisions to be less than $0.2$~quanta/s at low temperature.  At the highest trap temperatures, the LT stage temperature and the ion lifetimes are approximately equal to that achieved in the LT stage heating experiments, so we expect a contribution to the heating rate due to collisions no larger than $\gamma_{e}\sim\nobreak10$~quanta/s.   The Langevin collisions are typically over an order of magnitude less frequent than the elastic collisions (see Appendix~\ref{app_collisions}).  Since the Langevin rate is an upper bound, we can conservatively say, as a rule of thumb, that if $\tau_{L}/t_{D}\gg 100$ and $\tau_{L}\thinspace \frac{d\langle n\rangle}{dt}\gg 100$ are satisfied, background collisions will not be a significant cause of apparent heating~\footnote{This rule of thumb may have more limited validity in the case of ion species significantly less likely to chemically react with background constituents.}.  This is indeed the case for the temperature-dependent data presented above, without heating of the LT stage. Additionally, this conclusion is supported by the fact that our lowest heating rates are equivalent, when scaled by $d^{-4}$ for traps of different sizes, to the lowest heating rates measured in the field, including in systems with much longer ion lifetimes~\cite{MIT:HeatingvsT:PRL:08,brown2011coupled}.

If the heating rates measured here and elsewhere are limited by non-temperature-dependent sources of noise below a few tens of kelvin, as suggested by the low temperature plateaus observed elsewhere~\cite{MIT:HeatingvsT:PRL:08}, and hinted at here, true material-dependent noise mechanisms may continue to drop at lower temperatures.  This is reason to be optimistic that even lower heating rates can be obtained at temperatures probed here and accessible using widely available cryo-cooling technology if the limiting, and hopefully technical, noise sources can be eliminated.  Collision-induced effects should be negligible in most cryogenic systems under normal conditions, since ion lifetimes in these systems are very long, though care must be taken when measuring very low heating rates.  Sources of electric field noise may include nearby surfaces that remain at higher temperatures than the trap, especially in the case of the surface-electrode traps that will likely be required for large-scale quantum processing with trapped ions.  These traps are somewhat open and may provide less shielding at the ion location than traditional three-dimensional traps.  Voltage noise on electrodes from insufficiently filtered control electronics may also be a culprit if there is sufficient power near the trap frequency.  The lowest measured heating rates presented here correspond to voltage noise of  $\sim\nobreak1.1$~nV/$\sqrt{\rm Hz}$ for our trap geometry; after evaluation of our DAC control system and filtering, we expect less than $0.3$~nV/$\sqrt{\rm Hz}$ at the electrodes due to DAC electronics noise, excluding pickup.

Even though there is a possibility that the low-temperature heating rates measured here may not be ultimately limited by thermally activated material properties, they are still quite low; a $10$~$\mu$s two-qubit gate limited by the heating rates measured here would have an error rate of only $\sim\nobreak4\times10^{-5}$, below most estimates of the fault-tolerant threshold for quantum computation.  This is for ions $50$~$\mu$m from the electrode surface, suggesting that low error rates are possible for traps similar to this one in a large array.  In addition, it may be that combining low temperature electrodes with in situ surface cleaning can lead to even lower electric field noise levels than have been measured using either technique alone; this would potentially allow fault-tolerant error rates from ion motional heating for trap geometries significantly smaller, and hence potentially more scalable, than those explored here.

\begin{acknowledgments}
We thank Peter Murphy and Jacob Zwart for assistance with ion-trap-chip packaging.  This work is sponsored by the Assistant Secretary of Defense for Research \& Engineering under Air Force contract number FA8721-05-C-0002. Opinions, interpretations, conclusions, and recommendations are those of the authors and are not necessarily endorsed by the United States Government.
\end{acknowledgments}

\appendix

\section{Experimental Methods}
\label{app_methods}

To destabilize dark states in the $S_{1/2}$--$D_{3/2}$ manifold of the $^{88}$Sr$^{+}$ ion during Doppler cooling on the $S_{1/2}$--$P_{1/2}$ transition (near $422$~nm), and to reduce the frequency stability requirements, we use a diode laser near $1092$~nm broadened to an approximately $0.8$~GHz linewidth.  To maintain stability of the external-cavity diode laser used to address the narrow $S_{1/2}$--$D_{5/2}$ transition near 674~nm, we feed back to the diode current on a fast time scale, and to the grating angle on a slower time scale, via a Pound-Drever-Hall derived error signal from an ultra-stable ULE reference cavity.  This cavity is mounted vertically in a two-stage temperature-controlled vacuum system mounted on an active vibration isolation platform.  

Sideband cooling (SC) is performed on the $S_{1/2}$--$D_{5/2}$ transition in a pulsed manner~\cite{PhysRevLett.75.4011}. Optical pumping (OP) is required for state initialization to the $|S_{1/2},m_{j}=-1/2\rangle$ state and to compensate for depumping to the $|S_{1/2},m_{j}=1/2\rangle$ state during SC.  One cycle of SC consists of a pulse of light red-detuned by the trap frequency from the $|S_{1/2},m_{j}=-1/2\rangle$ $\rightarrow$ $|D_{5/2},m_{j}=-5/2\rangle$ transition, followed by a quench pulse of typically $1$~$\mu$s in length.  One cycle of OP consists of a pulse of light resonant with the $|S_{1/2},m_{j}=1/2\rangle$ $\rightarrow$ $|D_{5/2},m_{j}=-3/2\rangle$ transition, followed by a quench pulse.  We intersperse OP cycles during SC:  typically a few OP cycles are inserted after several SC cycles to form a cooling block, and we combine several cooling blocks in series.  The SC pulse lengths grow in each subsequent block to compensate for the reduced sideband Rabi frequency as the ion cools. Typical total times for Doppler cooling and SC to the ground state are $5$--$10$~ms.

\section{Calculation of Background Density from Motional Heating Rate}
\label{app_collisions}

Elastic collisions (rate $\gamma_{e}$) will lead to an apparent heating rate due to increased red sideband excitation in a small fraction of experiments averaged together with experiments without collisions for each heating rate measurement.  When using the sideband-ratio technique, this will appear as a sideband ratio~$r$ that is proportional to~$\gamma_{e}$.  For $\gamma_{e}t_{D}< 1$, where $t_{D}$ is the probe delay time, which dominates the experiment time for low heating rates, this will lead to an inferred average occupation due to collisions

\begin{equation}
\langle n\rangle=\frac{\gamma_{e}t_{D}}{1-\gamma_{e}t_{D}}.
\label{eq_nbar_coll}
\end{equation}

\noindent The ensemble of experiments in this case may not correspond to a thermal state, though the validity of the sideband-ratio technique requires this assumption.  The violation of this assumption will lead one to infer a steady heating rate.  

The combination of elastic collisions and heating from a more typical (perhaps anomalous) source at a rate $\gamma_{a}$ will lead to a sideband ratio of

\begin{equation}
 r=\gamma_{e}t_{D}+(1-\gamma_{e}t_{D})\frac{\gamma_{a}t_{D}}{1+\gamma_{a}t_{D}},
\end{equation}

\noindent and therefore a measured average occupation number of

\begin{equation}
\langle n \rangle = \frac{\left(\gamma_{e}+\gamma_{a}\right)t_{D}}{1-\gamma_{e}t_{D}}.
\label{eq_nbar_both}
\end{equation}

\noindent Due to this nonlinear relation, apparent heating due to collisions will look nonlinear in $t_{D}$ for appreciable values of $\gamma_{e}t_{D}$.  For $\gamma_{e}t_{D}\ll 1$, where our measurements are typically made, Eq.~\ref{eq_nbar_both} gives a linear relation in $t_{D}$, and so the effect of rare collisions will be the appearance of an additional steady heating rate $\gamma_{e}$.  We see an apparent (combined) heating rate of $32$~quanta/s with LT stage heating.  A fit to the heating rate data using Eq.~\ref{eq_nbar_both} (plus a zero-time occupation to account for imperfect preparation in the ground state) yields $\gamma_{e}=10(1)$~s$^{-1}$, assuming $13$~quanta/s due to anomalous heating (as measured without LT stage heating).

The elastic collision rate is proportional to the local density $n_{d}$ of the background gas constituents, and it depends on the temperature and polarizability of these constituents through the cross-section $\sigma$ and relative velocity $\tilde{v}\equiv\left( 2k_{B}T/\mu\right)^{1/2}$, with $\mu$ the reduced mass, as~\cite{NIST:ExpIssueswithIons:JresNIST:98}

\begin{equation}
\gamma_{e}= n_{d} \langle \sigma v \rangle \approx 1.23\times 10^{5}\medspace n_{d}\thinspace \tilde{v}^{1/3}\thinspace \alpha^{2/3}
\end{equation}

\noindent for a neutral-polarizing interaction between the ion and a neutral.  Here the velocity $v$ has been averaged over a thermal distribution.  For the background molecules expected here (predominantly H$_{2}$, N$_{2}$, and O$_{2}$ for temperatures $55$~K and below), we can use the apparent measured heating rate to calculate a background density of approximately $n_{d}=1.0(1)\times10^{9}$~cm$^{-3}$, equivalent to within $6$\% for the different background species considered here.  The calculated density is not strongly dependent on assumed molecular temperature ($\propto T^{1/6}$) or species mass, due to the weak dependence on the velocity.  The density is of course very strongly dependent on the temperatures of the surfaces in the vacuum system, however, due to the nature of cryoadsorption.

While an increased elastic collision rate due to the higher background gas density present during the $T=16$~K LT-stage temperature experiments leads to apparent average ion heating, the accompanying increased inelastic collision rate is likely the primary cause of the reduced ion lifetime $\tau_{L}$ under these conditions.  Using the background density measured above, we can estimate the inelastic collision rate $\gamma_{i}$.  This rate is lower than the elastic collision rate, since neutrals must penetrate the angular momentum barrier to chemically interact with the ion.  We can estimate it from the Langevin rate as~\cite{NIST:ExpIssueswithIons:JresNIST:98}

\begin{equation}
\gamma_{i}=n_{d}\thinspace \sigma v = n_{d}\thinspace q \sqrt{\frac{\pi \alpha}{\epsilon_{0} \mu}}
\end{equation}

\noindent for an ion of charge $q$ and neutral species of speed $v$ (the ion is assumed to be at rest).  For the density measured above, we find a Langevin rate of $0.6$~s$^{-1}$ to $1.6$~s$^{-1}$.

\bibliography{jc_bib1,IonLoading_PRAfinal}

\end{document}